\documentclass[preprint,aps,prd,showpacs,superscriptaddress,nofootinbib,tightenlines]{revtex4}

\usepackage{amsfonts}
\usepackage{multirow}
\usepackage{mathrsfs}
\usepackage{graphicx}
\usepackage{amsmath}
\usepackage{amssymb}
\usepackage{bm}
\usepackage{bbm}
\usepackage{color}
\usepackage{array}
\usepackage{diagbox}
\usepackage{float}
\usepackage{makecell}
\allowdisplaybreaks[4]

\def\OMIT#1{}

\def\hlinew#1{%
  \noalign{\ifnum0=`}\fi\hrule \@height #1 \futurelet
   \reserved@a\@xhline}
\usepackage{array}
\newcommand{\PreserveBackslash}[1]{\let\temp=\\#1\let\\=\temp}
\newcolumntype{C}[1]{>{\PreserveBackslash\centering}p{#1}}
\newcolumntype{R}[1]{>{\PreserveBackslash\raggedleft}p{#1}}
\newcolumntype{L}[1]{>{\PreserveBackslash\raggedright}p{#1}}

\newcommand{\nn}{\nonumber}

\newcommand{\beq}{\begin{equation}}
\newcommand{\eeq}{\end{equation}}
\newcommand{\bqa}{\begin{eqnarray}}
\newcommand{\eqa}{\end{eqnarray}}

\newcommand{\jpsi}{J/\psi}

\newcommand\fverb{\setbox\fverbbox=\hbox\bgroup\verb}
\newcommand\fverbdo{\egroup\medskip\noindent%
			\fbox{\unhbox\fverbbox}\ }
\newcommand\fverbit{\egroup\item[\fbox{\unhbox\fverbbox}]}
\newbox\fverbbox

\makeatletter

\newcommand{\Rmnum}[1]{\expandafter\@slowromancap\romannumeral #1@}
\makeatother

\begin{document}
\title{\mbox{}\\[10pt]
Revisiting the line shape of $e^+e^-\to \jpsi \eta$ cross section}
\author{Chang-Man Gan~\footnote{cm.gan@outlook.com}}
\affiliation{School of Physical Science and Technology, Southwest University, Chongqing 400700, China\vspace{0.2cm}}
\author{Jun-Kang He~\footnote{hejk@hbnu.edu.cn}}
\affiliation{College of Physics and Electronic Science, Hubei Normal University, Huangshi 435002, China\vspace{0.2cm}}
\affiliation{Key Laboratory of Quark and Lepton Physics (MOE), Central China
Normal University, Wuhan 430079,China\vspace{0.2cm}}
\author{Wen-Long Sang~\footnote{wlsang@swu.edu.cn}}
\affiliation{School of Physical Science and Technology, Southwest University, Chongqing 400700, China\vspace{0.2cm}}
\author{Min-Zhen Zhou~\footnote{zhoumz@swu.edu.cn}}
\affiliation{School of Physical Science and Technology, Southwest University, Chongqing 400700, China\vspace{0.2cm}}

\date{\today}
\begin{abstract}
We calculate the cross sections for the processes $e^+e^-\to \jpsi \eta$ and $e^+e^-\to \jpsi \eta^\prime$ at various CM energies $\sqrt{s}$.
We first predict these cross sections by combining NRQCD with LC factorization.  The predicted cross sections are on the order of several femtobarns for 
$e^+e^-\to \jpsi \eta^\prime$, and less than 1 fb for $e^+e^-\to \jpsi \eta$,  which are significantly smaller than the experimental measurements. It is anticipated that the cross sections are dominated 
by resonant contributions when $\sqrt{s}$ is close to the resonance mass. In this study, we employ the Vector Meson Dominance (VMD) model to predict these resonant contributions. The effective coupling constants between the photon and the resonance, as well as between the resonance and $J/\psi\eta$ are extracted from the data either provided by the latest PDG or predicted by theoretical calculations.
Taking the predictions from the factorization calculation as the continuum contribution, we predict the cross section of $e^+e^-\to \jpsi \eta$ through a coherent sum of contributions from various resonances and the continuum. The relative phase angles between these contributions are determined through a least-$\chi^2$ fit to the experimental data. 
We then compare our theoretical predictions with the experimental data. Additionally, we find our theoretical prediction for the cross section of $e^+e^-\to \jpsi \eta$ is significantly larger than those for $e^+e^-\to \jpsi \eta^\prime$ measured by the BESIII collaboration.  

\end{abstract}
\maketitle

\section{introduction}
Over the past two decades, several vector charmonium and charmonium-like states have been investigated through the processes  $e^+e^-\to \jpsi \eta$ and  $e^+e^-\to \jpsi \eta^\prime$. By analyzing the cross-section line shapes of these processes, the resonance structures, including their masses and widths, can be determined. Furthermore, experimental data provide valuable insights into the hadronic transitions (either an $\eta$ or $\eta^\prime$) between these states and $\jpsi$,  thereby aiding in elucidating the nature of these states.

The initial search for the process $e^+e^-\to \jpsi \eta$  was conducted by the CLEO collaboration~\cite{CLEO:2006ike}. Subsequently, the BESIII collaboration first measured the cross section of this process at a center-of-mass (CM) energy of $\sqrt{s}=4.009$ GeV~\cite{BESIII:2012fdg}. Later, the BELLE collaboration measured the cross section for $e^+e^-\to \jpsi \eta$ via initial state radiation~\cite{Wang:2012bgc}. Two resonance structures have been identified to account for the $\jpsi\eta$ invariant mass distribution. In 2020, the BESIII collaboration measured the cross section over a range of CM energies between $3.81$ and $4.60$ GeV, identifying three resonant states through a maximum-likelihood fit~\cite{BESIII:2020bgb}. With additional data samples, the BESIII collaboration further refined their measurements in Ref.~\cite{BESIII:2023tll}.  Additionally, searches for the $\psi(4S)$ resonance through $e^+e^- \to J/\psi\eta$ based on Belle measurements have been conducted, but no significant signal was observed in the invariant mass spectra~\cite{Gao:2015yoi}.

In the meantime, the cross section of the process $e^+e^-\to \jpsi \eta^\prime$ was measured by BESIII collaboration~\cite{BESIII:2016vio,BESIII:2019nmu}. The line shape of this process can be reasonably described by a coherent sum of two resonant states. It is worth noting that the cross section of $e^+e^-\to \jpsi \eta^\prime$ is found to be an order of magnitude lower than that of $e^+e^-\to \jpsi \eta$.

Extensive theoretical studies have been conducted on the hadronic transitions of vector charmonium and charmonium-like states to $\jpsi$ via $\eta$ or $\eta^\prime$, as well as on the cross-section line shapes of $e^+e^-\to \jpsi \eta$ and $e^+e^-\to \jpsi \eta^\prime$. 

In Ref.~\cite{Casalbuoni:1992fd}, the transition $\psi(2S) \to \jpsi \eta$ was studied using an effective chiral Lagrangian for charmonium states and light pseudoscalars. Refs.~\cite{Guo:2009wr,Guo:2010ak} investigated the impact of intermediate charmed meson loops on $\psi(2S)\to J/\psi\eta$, demonstrating that these loops significantly alter the decay amplitude. Ref.~\cite{Chen:2012nva} examined the $\eta$  transition process between  $\psi(4040)/\psi(4160)$  and  $J/\psi$,
as well as the $\eta^\prime$ transition between $\psi(4160)$ and $J/\psi$ by incorporating charmed meson loops within an effective Lagrangian approach to enhance the decay amplitudes. 
The upper limit of the branching ratio for  $\psi(4S)\to J/\psi\eta$ was set in Ref.~\cite{Chen:2014sra} using a similar framework. Refs.~\cite{Anwar:2016mxo,Anwar:2017urj} studied the hadronic transitions of higher vector charmonia with the emission of an $\eta$ meson, using a model that describes the creation of a light meson in heavy quarkonium transitions. Additionally, hadronic transitions have been explored based on multipole expansion~\cite{Gottfried:1977gp,Voloshin:1978hc,Yan:1980uh,Kuang:2006me}.

In Ref.~\cite{Wang:2011yh}, the open charm effects in $e^+e^-\to \jpsi \eta$ was examined, revealing that the cross-section line shape is strongly influenced by these effects. Ref.~\cite{Qiao:2014pfa} computed the cross sections of $e^+e^-\to \jpsi \eta$ and $e^+e^-\to \jpsi \eta^\prime$ by combining the nonrelativistic QCD (NRQCD) framework with the light-cone (LC) approach. The theoretical cross section for $e^+e^-\to \jpsi \eta$ was found to be comparable with the experimental data. Additionally, the authors considered resonance state effects by introducing effective coupling constants between $\psi$ and $c\bar{c}$, as well as $\psi$ and $e^+e^-$. With these coupling constants set to specific values, and summing the non-resonance as well as resonance contributions, their theoretical predictions were able to well explain the experimental data. In Ref.~\cite{Zhang:2018gwe}, three resonant structures were identified in the cross-section line shape of $e^+e^-\to \jpsi \eta$. The resonance parameters for these structures were determined by fitting the cross section data using a least-$\chi^2$ method, and the branching fractions of these resonances to $\jpsi\eta$ were also present. 
In Ref.~\cite{Peng:2024xui}, five theoretically constructed charmonia in the mass range of $4.0–4.5$ GeV
were employed to perform a combined fit to the experimental data. The branching ratios for the decay of these charmonium states into $\jpsi\eta$  via the hadronic loop mechanism were used as input parameters. 

As reported in Refs.~\cite{BESIII:2016vio,BESIII:2019nmu}, the measured cross section for  $e^+e^-\to \jpsi \eta^\prime$  is approximately an order of magnitude lower than that for $e^+e^-\to \jpsi \eta$. This discrepancy is inconsistent with the theoretical predictions presented in Ref.~\cite{Qiao:2014pfa}. Moreover, it is somewhat perplexing that the non-resonant contribution to the cross section of $e^+e^-\to \jpsi \eta$ is significantly larger than that of $e^+e^-\to \jpsi \eta_c$~\cite{Belle:2002tfa,Braaten:2002fi,Liu:2002wq}, given that the former is a loop-induced process.
Motivated by these puzzles, we revisit the processes of $e^+e^-\to \jpsi \eta$ and  $e^+e^-\to \jpsi \eta^\prime$ in thi work. We will compute the contributions from non-resonant states using the factorization approach. Specifically, $\jpsi$ will be treated within the framework of NRQCD, while $\eta$ and $\eta^\prime$ will be treated using the LC approach. 
We will account for the contributions from both the quark-antiquark and gluonium components~\footnote{It is worth noting that the quantum numbers, mass, production, and decay properties of the X(2370) particle have been found to be consistent with the features of a glueball~\cite{BESIII:2023wfi}, which has long been the subject of intensive experimental searches. This discovery provides strong experimental evidence supporting the existence of glueballs.}. Additionally, the contributions from resonant states will be computed using the VMD model. 

The remainder of this paper is organized as follows. In Sec.~\ref{sec:FF}, we define the electromagnetic form factor and derive the corresponding cross section. In Sec.~\ref{sec-factorization-framework}, we outline the approach for computing the non-resonant contributions. In Sec.~\ref{sec-resonance}, we present the methodology for calculating the resonant contributions using the VMD model. Our phenomenological predictions and discussions are provided in Sec.~\ref{sec-phenomenology}. Finally, we summarize our results in Sec.~\ref{sec-summary}.

\section{The general formulas~\label{sec:FF}}

The amplitude for the process $e^+(k_1) + e^-(k_2) \rightarrow J/ \psi(P_1) + H(P_2)$, where $H$ denotes either $\eta$ or $\eta^\prime$, can be written as
\begin{eqnarray}\label{amp-all}
	- i \mathcal{M}= -\frac{i}{s}L_{\mu} \mathcal{A}^{\mu},
\end{eqnarray}
where $e$ is the elementary charge, $k_1$, $k_2$, $P_1$, $P_2$ represent the momenta of $e^+$, $e^-$, $J/\psi$, $H$, respectively. $s=(P_1 + P_2)^2$ is the square of the CM energy. The leptonic current  $L_{\mu}$ is given by
\begin{eqnarray}\label{leptonic-current}
	L_{\mu}=e\, \bar{v} \left(k_1\right) \gamma_{\mu} u\left(k_2\right).
\end{eqnarray}
Here, $ \mathcal{A}^{\mu}$ represents the amplitude of $\gamma^* \to J/\psi H$, which can be expressed in terms of the time-like electromagnetic form factor $F(s)$~\cite{Braaten:2002fi}:
\begin{eqnarray}\label{em-form-factor}
	\mathcal{A}^{\mu} = e F(s) \epsilon^{\mu \nu \rho \sigma} P_{1\nu} P_{2\rho} \epsilon^*_{\sigma},
\end{eqnarray}
where $\epsilon$ represents the polarization vector of $J/\psi$. The tensor structure in this expression is determined by parity conservation and Lorentz invariance.  The cross section for the process $e^+ e^- \to J/\psi H$  can then be expressed as 
\begin{eqnarray}\label{cross-section-re}
	\sigma(e^+ e^- \to J/\psi H) = \frac{4\pi \alpha^2}{3} \left( \frac{|\boldsymbol{P_{1}}|}{\sqrt{s}} \right)^3 |F(s)|^2,
\end{eqnarray}
where $\alpha$ is the electromagnetic coupling constant,  $|\boldsymbol{P_{1}}|$ is the magnitude of the three-momentum of the $\jpsi$  in the CM frame, given by
\begin{eqnarray}\label{eq:pcm}
|\boldsymbol{P_{1}}|=\frac{\lambda^{1/2}(s,M_{\jpsi}^2,M^2_H)}{2\sqrt{s}}
\end{eqnarray}
with $M_H$ denotes the mass of $H$, and the K\"allen function is defined via $\lambda(x,y,z)=x^2+y^2+z^2-2xy-2yz-2xz$. 

Thus, to determine the cross section, it suffices to compute the form factor  $F(s)$. In the subsequent sections, we will calculate this form factor, taking into account both non-resonant and resonant contributions.

\section{Non-resonant contribution~\label{sec-factorization-framework}}

\begin{figure}[htbp]
	\centering
	\includegraphics[width=0.8 \textwidth]{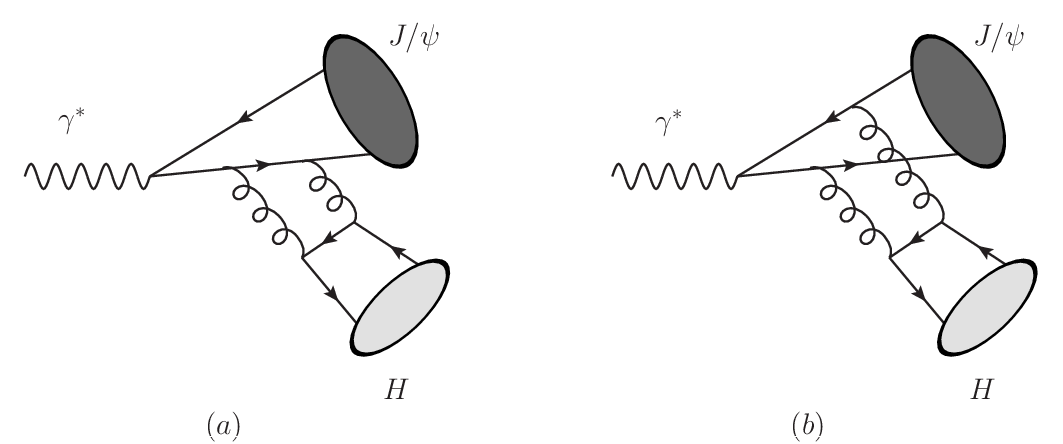}
	\caption{Typical Feynman diagrams for the process $\gamma^{*} \to J/\psi H$  with contributions from the quark-antiquark component of $H$.
		\label{fig-feynman-diagram-quark}}
\end{figure}

\begin{figure}[htbp]
	\centering
	\includegraphics[width=0.8\textwidth]{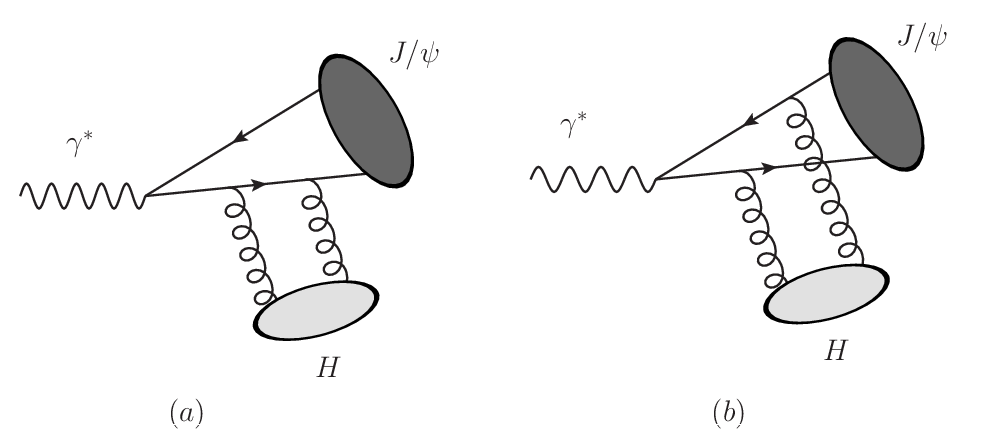}
	\caption{
 Typical Feynman diagrams for the process $\gamma^{*} \to J/\psi H$  with contributions from the gluonium component of $H$. \label{fig-feynman-diagram-gluon}}
\end{figure}

We can employ factorization to compute the form factor $F(s)$. Specifically, we use NRQCD to handle the production of the non-relativistic meson $J/\psi$, while adopting the LC approach for the production of the light meson $H$. The typical Feynman diagrams for the process $\gamma^* \rightarrow \jpsi H$ are illustrated in Fig.~\ref{fig-feynman-diagram-quark} and Fig.~\ref{fig-feynman-diagram-gluon}. 


Using the NRQCD factorization~\cite{Bodwin:1994jh}, we can express the form factor $F(s)$ to the lowest order in velocity as
\begin{eqnarray}\label{nrqcd-formula}
		F(s) &=& \mathcal{C}(H)\times \frac{\sqrt{2M_{J/\psi}} \langle J/\psi | \chi^\dagger \boldsymbol{\sigma} \cdot \boldsymbol{\epsilon} \psi | 0 \rangle}{\sqrt{2N_c}2m_c}, 
\end{eqnarray}
where the non-relativistically normalized NRQCD matrix elements can be related to the radial Schr\"{o}dinger wave function of $J/\psi$ at the origin via
\beq
\langle J/\psi | \chi^\dagger \boldsymbol{\sigma} \cdot \boldsymbol{\epsilon} \psi | 0 \rangle \approx \sqrt{\frac{3}{2\pi}} \big|R_{J/\psi}(0)\big|.
\eeq
The factor $\mathcal{C}(H)$, which still contains nonperturbative effects for $H$, can be further factorized using LC factorization~\cite{Lepage:1979zb,Lepage:1979za,Lepage:1980fj,Efremov:1978rn,Efremov:1979qk,Duncan:1979ny,Duncan:1979hi} (for a review~\cite{Chernyak:1983ej}). 

Before applying the factorization, it is crucial to analyze the structures of $\eta$ and $\eta^\prime$. It is convenient to use the quark-flavor (QF) bases, where the basis vectors are $\eta_{q}= (u \bar{u} +d \bar{d})/\sqrt{2}$ and 
$\eta_{s} = s \bar{s}$. 
The states can then be decomposed as~\cite{Feldmann:1998vh}
\begin{eqnarray}\label{QF-bases}
	& | \eta \rangle = \cos \phi  | \eta_{q} \rangle - \sin \phi  | \eta_{s} \rangle, \\
	& | \eta^{\prime} \rangle = \sin \phi  | \eta_{q} \rangle + \cos \phi | \eta_{s} \rangle.
\end{eqnarray}

We can now employ LC factorization to express
\begin{eqnarray}\label{LC-factorization}
\mathcal{C}(H) =  \sum_{i=u, d, s, g}f^i_{H}\int_0^1 dx \, c_{H}^i(x, \mu) \phi_{H}^i(x, \mu),
\end{eqnarray}
where $f_{H}^u$, $f_{H}^d$ and $f_{H}^s$ are effective decay constants. And these constants can be expressed in terms of three phenomennological parameters $f_q$, $f_s$ and the mixing angle $\phi$~\cite{Feldmann:1998vh}
\begin{eqnarray}\label{decay-constants-qs}
	\notag
	f_{\eta}^{u(d)} &=f_{q}\cos \phi/\sqrt{2} , \quad f_{\eta}^{s}&=-f_{s} \sin \phi, \\
	f_{\eta^{\prime}}^{u(d)} &=f_{q} \sin \phi/\sqrt{2}, \quad f_{\eta^{\prime}}^{s}&=f_{s} \cos \phi .
\end{eqnarray}
Additionally, $f_{H}^g = (f_{H}^u + f_{H}^d + f_{H}^s)/\sqrt{3}$ represents the effective decay constant of the gluonium component.
Here $c_{H}^i(x, \mu)$ and $\phi_{H}^i(x, \mu)$ represent hard-scattering coefficients and LC distribution amplitudes, respectively.
All nonperturbative effects in $\mathcal{C}(H)$ are encoded in $f_{H}^i$ and $\phi_{H}^i(x, \mu)$, while
the hard-scattering coefficients $c_{H}^i(x, \mu)$ are fully perturbative and therefore can be calculated using perturbative QCD.  

The LC distribution amplitudes $\phi_{H}^q$ are defined by~\cite{Kroll:2002nt,Ball:2007hb}
\begin{eqnarray}\label{eta-matrix}
	\left.\langle H (P_2) \right.|\bar{q}_{\alpha}(z) [z,-z] q_{\beta}(-z) \left. |0 \right. \rangle =\frac{i}{4} f^{q}_{H} (P\!\!\!\slash_2 \gamma^{5})_{\beta \alpha}  \int_{0}^{1} \mathrm{d} x e^{i (2x-1) P_2 \cdot z} \phi^{q}_{H}(x),
\end{eqnarray}
where $q$ denotes either the $u$ quark or the $d$ quark, and $\phi_{H}^g$ is defined by~\cite{Kroll:2002nt,Ball:2007hb}~\footnote{The path-ordered gauge factor $[z,-z]$ has been dropped out from the matrix elements.}
\begin{eqnarray}\label{eta-gluonic}
	\left\langle H(P_2)\left|A_{\alpha}^{a}(z) A_{\beta}^{b}(-z)\right| 0\right\rangle =\frac{1}{4} \epsilon_{\alpha \beta \rho \sigma} \frac{\bar{P_2}^{\rho} P_{2}^{\sigma}}{\bar{P_2} \cdot P_2} \frac{C_{F}}{\sqrt{3}} \frac{\delta^{a b}}{8} f_{H}^{g} \int_{0}^{1} \mathrm{d} x e^{i(2x-1) P_2 \cdot z} \frac{\phi^{g}_{H}(x)}{x(1-x)}
\end{eqnarray}
with the vector $\bar{P_2}^{\rho}=(P^{0}_{2},-\boldsymbol{P_{2}})$.

We briefly describe our perturbative calculation for the hard-scattering coefficients. We use the FeynArts package~\cite{Hahn:2000kx} to generate partonic-level Feynman diagrams and amplitudes for the processes $\gamma^* \to c(P_1/2)\bar{c}(P_1/2) +q(xP_2)\bar{q}(\bar{x}P_2)$ and $\gamma^* \to c(P_1/2)\bar{c}(P_1/2) +g(xP_2)g(\bar{x}P_2)$.  
We employ the following projectors to extract the intended spin and color quantum numbers.
The projectors are~\cite{Petrelli:1997ge}
\begin{eqnarray}\label{c-proj}
	\Pi_{1}= \frac{1}{\sqrt{2N_c}} \epsilon\!\!\slash^{\ast}  \left(\frac{P\!\!\!\!\slash_{1}}{2} +m_c\right),
\end{eqnarray}
 for $c\bar{c}$,
\begin{eqnarray}\label{q-proj}
	\Pi_2^{q }= \frac{1} {4 N_c}\left(i P\!\!\!\!\slash_{2} \gamma_{5}\right),
\end{eqnarray}
for $q\bar{q}$,
and
\begin{eqnarray}\label{g-proj}
	\left(\Pi_2^{g}\right)^{ab}_{\alpha\beta}= \frac{1}{4} \epsilon_{\alpha \beta \rho \sigma} \frac{\bar{P_2}^{\rho} P_2^{\sigma}}{\bar{P_2} \cdot P_2} \frac{C_{F}}{\sqrt{3}} \frac{\delta^{a b}}{8} \frac{1}{x(1-x)},
\end{eqnarray}
for $gg$. In Eq.~(\ref{g-proj}),
$a, b$ and $\alpha, \beta$  denote the color indices and Lorentz indices of the two gluons, respectively.
These projectors allow us to extract the spin-triplet S-wave component of $c\bar{c}$
and the spin-singlet S-wave component of $q\bar{q}$ and $gg$. 

We then use the FeynCalc/FormLink package~\cite{Mertig:1990an,Feng:2012tk}
to perform Dirac traces and Lorentz contractions.  The
Apart~\cite{Feng:2012iq} and FIRE~\cite{Smirnov:2019qkx} are employed for partial fraction decomposition and integration-by-parts reduction. The one-loop master integrals (MIs) are analytically computed using PackageX~\cite{Patel:2015tea}. Finally, using the definition of the form factor in Eq.~\eqref{em-form-factor}, we determine the hard-scattering coefficients $c^i_H$.

\section{Resonant contribution~\label{sec-resonance}}

\begin{figure}[htbp]
	\centering
	\includegraphics[width=0.55\textwidth]{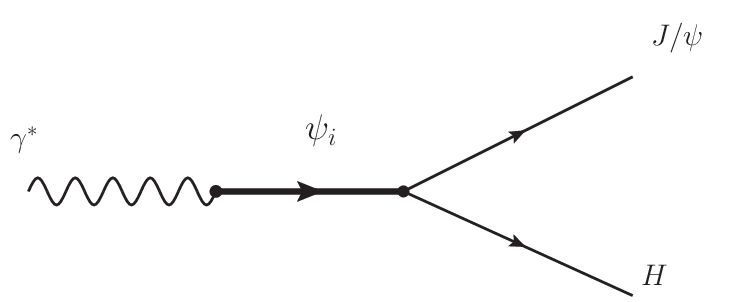}
	\caption{Schematic diagram for the process $\gamma^{*} \to J/\psi  H$ based on VMD Model.
		\label{fig-feynman-diagram-VMD}}
\end{figure}

In this section, we outline the theoretical framework for computing the form factor $F(s)$ from resonant contributions. We utilize the VMD model~\cite{Bauer:1977iq}, an effective approach for describing the resemblance of photon and hadron interactions. Within this model, the photon is assumed to couple to hadrons primarily via its conversion into a virtual vector meson, which subsequently interacts with the hadrons. A schematic diagram the of process $\gamma^*\to J/\psi  H$ is shown in Fig.~\ref{fig-feynman-diagram-VMD}. By introducing intermediate vector mesons, the VMD model provides a simplified yet robust method to describe photon-hadron interactions  in the energy region near the resonance, thereby circumventing the complexities associated with direct QCD interactions.
 
To describe these interactions,
it is essential to introduce the effective Lagrangians for both the $ \gamma V$ and $V J/\psi H$ couplings. Here, $V$ represents the charmonium vector resonance $\psi_i$. 

The $ \gamma V$ effective coupling is described by the following Lagrangian~\cite{Bauer:1977iq,Zhao:2010mm,Zhao:2006gw}:
\begin{eqnarray}\label{L-rV}
\mathcal{L}_{\gamma V} = i  {g_{\gamma V} M_V^2} V^\mu A_\mu,
\end{eqnarray}
where  $V^\mu $ and $A^\mu$ denote the $V$ filed and the photon field, respectively. $M_V$ is the mass of the vector meson $V$, and $g_{\gamma V}$ is the dimensionless coupling constant between the photon and the vector meson $V$. The constant $g_{\gamma V}$ can be determined from the decay process $V\to e^+e^-$ via the relation
\begin{eqnarray}\label{width-Vee}
   g_{\gamma V}=\left[\frac{3\, \Gamma(V \to e^{+} e^{-})}{ M_V\alpha}\right]^{1 / 2},
\end{eqnarray}
where $ \Gamma(V \to e^{+} e^{-})$ denotes the decay width of the process $V \to e^{+} e^{-}$. 

The $V J/\psi H$ effective coupling is described by 
the following Lagrangian~\cite{Zhao:2010mm,Zhao:2006gw,Intemann:1983yj,Lichard:2010ap}:
\begin{eqnarray}\label{L-VVP}
\mathcal{L}_{V J/\psi H} =i g_{VJ/\psi H}\mathcal{F}(|\boldsymbol{P_{1}}|^2) \epsilon^{\mu\nu\alpha\beta} \partial_\mu V_\nu \partial_\alpha V^{\prime}_\beta P,
\end{eqnarray}
where ${\boldsymbol P_1}$ represents the three momentum of the $J/\psi$,  $g_{V J/\psi H}$ is the corresponding coupling constant, $V^{\prime}_\beta$ represents the $J/\psi$ filed, and $P$ represents the $H$ field. The interaction in Eq.~(\ref{L-VVP}) satisfies Lorentz invariance and parity conservation.

Here, $\mathcal{F}(|\boldsymbol{P_{1}}|^2)$ is a model-dependent form factor, which can be interpreted as a measure of the overlap of the wave function. In practical calculations, this form factor is typically employed in an empirical form~\cite{Godfrey:1985xj,Amsler:1995td,Zhao:2010mm}:
\begin{eqnarray}\label{FF2}
  \mathcal{F}(|\boldsymbol{P_{1}}|^2) = e^{-|\boldsymbol{P_{1}}|^2/ 8\beta^2},
\end{eqnarray}
where the parameter $\beta$ lies in the range of 300-500 MeV for the charmonia decays.  

The decay width of $V\to J/\psi H$ can be derived using Eq.~(\ref{L-VVP}). Thus, the coupling constant $g_{V J/\psi H}$ can be determined from the decay width of $V\to J/\psi H$ via the relation 
 \begin{eqnarray}\label{gvvp-m}
    g_{V  J/\psi H}=\left[\frac{12 \pi  \Gamma(V \to  J/\psi H)}{\left|\boldsymbol{P}\right|^{3} \mathcal{F}^2(|\boldsymbol{P}|^2)}\right]^{1 / 2},
\end{eqnarray}
where $\boldsymbol{P}$ is the three momentum of $J/\psi$ in the rest frame of $V$.

The amplitude for the process $\gamma^{\ast}\to J/\psi  H$ in the VMD model can be expressed as follows:
\begin{eqnarray}\label{vmd-amp}
     \mathcal{A}_{\rm VMD}^{\mu} = g_{\gamma V}M_V^2\frac{1}{s - M_V^2 + i M_V \Gamma_V} g_{VJ/\psi H}\mathcal{F}(|\boldsymbol{P_{1}}|^{2}) \epsilon^{\mu\nu\alpha\beta} \epsilon_\nu^{\ast} P_{1\alpha} P_{2\beta},
\end{eqnarray}
where $\Gamma_V$  represents the total width of the intermediate vector meson $V$. The factor $ (s - M_V^2 + i M_V \Gamma_V)^{-1}$  accounts for the propagator of the intermediate resonance $V$.

By comparing Eq.~(\ref{em-form-factor}) with Eq.~(\ref{vmd-amp}), we can derive the expression for the form factor $F(s)$ in the VMD model, mediated by the intermediate state $V$
\begin{eqnarray}\label{vmd-FFs}
    F^{\rm VMD}_V(s) = \frac{g_{\gamma V} M_V^2}{e} \frac{1}{s - M_V^2 + i M_V \Gamma_V} g_{VJ/\psi H}\mathcal{F}(|\boldsymbol{P_{1}}|^{2}).
\end{eqnarray}

\section{Phenomenological results and discussions ~\label{sec-phenomenology}}

In our numerical calculations, we use the following values from the Particle Data Group (PDG)~\cite{ParticleDataGroup:2024cfk}:  $M_{\jpsi}= 3.097$ GeV, $M_{\eta}=0.548$ GeV, and $M_{\eta^{\prime}}=0.958$ GeV. The electromagnetic coupling constant is set to $\alpha=1/127$. 
\subsection{Non-resonant contribution}

We take the NRQCD matrix element to be
\bqa
\left| \langle J/\psi | \chi^\dagger \boldsymbol{\sigma} \cdot \boldsymbol{\epsilon} \psi | 0 \rangle \right|^{2}  \approx \frac{N_c}{2 \pi} R_{J / \psi}^{2}(0)= \frac{N_c}{2 \pi}\cdot 0.81 \; \text{GeV}^3,
\eqa
where the the radial Schr\"{o}dinger wave function of $J/\psi$ at the origin is evaluated using Buchm\"{u}ller-Type potential model~\cite{Buchmuller:1980su,Eichten:1995ch}. We set the running strong coupling constant to $\alpha_s(m_c)=0.31$, and the charm quark mass to $m_c=1.5$ GeV.

The decay constants $f_{q}$, $f_{s}$, and the mixing angle $\phi$ have been determined in the Feldmann-Kroll-Stech scheme through a fit to experimental data~\cite{Feldmann:1998vh}:
\begin{align}\label{coulpings1-qs}
	\notag
	f_{q} &=(1.07 \pm 0.02) f_{\pi}, \\
	f_{s} &=(1.34 \pm 0.06) f_{\pi}, \\ 	\notag
	\phi &= 39.3^{o} \pm 1.4^{o},
\end{align}
where the pion decay constant is $f_{\pi}=0.130$ GeV~\cite{Feldmann:1999uf}. 

Additionally, we adopt the LC distribution amplitudes $\phi^q_{H}(x,\mu)$ from Ref.~\cite{Agaev:2014wna}
\begin{eqnarray}\label{DA-q}
	\phi^q_{H}(x,\mu)=6x(1-x)\bigg[1 + \sum_{n=2,4,\cdots} d^q_{n}(\mu) C^{\frac{3}{2}}_{n}(2x-1)\bigg],
\end{eqnarray}
and $\phi^g_{H}(x,\mu)$ from Refs.~\cite{Ball:2007hb,Agaev:2014wna,Alte:2015dpo}
\begin{eqnarray}\label{DA-g}
	\phi^{g}_{H}(x,\mu) =30x^2(1-x)^2 \sum_{n=2,4,\cdots} d^{g}_{n}(\mu) C^{\frac{5}{2}}_{n-1}(2x-1),
\end{eqnarray}
where $d^{q}_{n}(\mu)$ and $d^g_n(\mu)$ are the Gegenbauer moments. 
In our numerical computation, we truncate the sum to $n=4$ in (\ref{DA-q}) and $n=2$ in (\ref{DA-g}).
The values of $d^q_2$, $d^q_4$, and $d^g_2$ are evaluated using different models~\cite{Agaev:2014wna}.
It is observed that our numerical result for the form factor $F(s)$ are insensitive to the choice of different models (see another example in Ref.~\cite{He:2019mpy}). Therefore, for concrete calculations, we adopt the values $d^q_2=0.20$, $d^q_4=0$, and $d^g_2=-0.31$ from Ref.~\cite{Agaev:2014wna}.

With all the necessary ingredients in hand, we proceed to make our numerical predictions. The theoretical predictions for the form factor $F(s)$ and the cross sections are summarized in Table~\ref{tab:C-S-eta} for the process $e^+ e^- \rightarrow J/\psi \eta$, and in Table~\ref{tab:C-S-etap} for the process $e^+ e^- \rightarrow J/\psi \eta^\prime$.
\begin{table}[!htbp]\small
	\caption{ The cross sections $\sigma$ (in units of $\times 10^{-4}$ pb) and the form factors $F(s)$ (in units of $\times 10^{-4}{\rm GeV}^{-1}$) of $e^+ e^- \rightarrow J/\psi \eta$ over the CM energy range from 3.65 to 10 GeV. $\sigma^q$, $\sigma^g$ and $\sigma$ represent the cross sections arising from the sole quark-antiquark component, the sole gluonium component, and the total contribution, respectively.}
	\label{tab:C-S-eta}
	\centering
	\setlength{\tabcolsep}{8pt}
	\renewcommand{\arraystretch}{1.6}
	\begin{tabular}{c c c c c c c c c c c}
		\hline\hline
		$\sqrt{s}$ & 3.65  & 3.7 & 3.8 & 3.9  & 4.0 &4.1 &4.2& 4.3&4.4 & 4.5\\
		\hline
		\multirow{2}*{$F(s)$} &23.1&19.1&13.9&10.6&8.4&6.8&5.6&4.7&4.0&3.5 \\
		&11.0$i$&9.9$i$&8.1$i$&6.8$i$&5.8$i$&4.9$i$&4.3$i$&3.7$i$&3.3$i$&2.9$i$\\
		\hline
		$\sigma^q$   &0.02  &0.6  &2.0&2.8&3.2&3.3 &3.2&3.0& 2.8&2.6\\
        \hline
        $\sigma^g$&0.005&0.1&0.2&0.1&0.1&0.07&0.05&0.04&0.03&0.02\\
        \hline
        $\sigma$&0.04&1.1&3.0&3.9&4.2&4.1&3.9&3.6&3.2&2.9\\
		\hline\hline
	\end{tabular}
	\begin{tabular}{c c c c c c c c c c c}
		\hline\hline
		$\sqrt{s}$& 4.6 & 4.7 & 4.8 & 4.9 & 5& 6 & 7 & 8 &9 & 10 \\
		\hline
		\multirow{2}*{$F(s)$} &3.0&2.6&2.3&2.1&1.8&0.7&0.4&0.2&0.1&0.08 \\
  &2.6$i$&2.3$i$&2.1$i$&1.9$i$&1.7$i$&0.7$i$&0.4$i$&0.2$i$&0.2$i$&0.1$i$ \\
		\hline
	$\sigma^q$ & 2.3 &2.1& 1.9& 1.7& 1.5&0.5& 0.2  &0.07& 0.03& 0.01 \\
        \hline
        $\sigma^g$&0.01&0.009&0.007&0.005&0.004&0.0003&0.00003&0.0&0.0&0.0\\
        \hline
        $\sigma$&2.6&2.3&2.0&1.8&1.6&0.5&0.2&0.07&0.03&0.01\\
		\hline\hline
	\end{tabular}

\end{table}

\begin{table}[!htbp]\small
	\caption{ The cross sections $\sigma$ (in units of $\times 10^{-4}$ pb) and the form factors $F(s)$ (in units of $\times 10^{-4}{\rm GeV}^{-1}$) of $e^+ e^- \rightarrow J/\psi  \eta^\prime$ over the CM energy range from 4.1 to 10 GeV. The symbols used have the same meaning as those in Table~\ref{tab:C-S-eta}.}
	\label{tab:C-S-etap}
	\centering
	\setlength{\tabcolsep}{8pt}
	\renewcommand{\arraystretch}{1.6}
	\begin{tabular}{c c c c c c c c}
		\hline\hline
		$\sqrt{s}$  & 4.1 & 4.2 & 4.3 & 4.4 & 4.5& 4.6 & 4.7 \\
		\hline
		\multirow{2}*{$F(s)$} &51.5&39.6&31.6&25.8&21.5&18.3&15.7\\
		  &+22.8$i$&+20.0$i$ &+17.6$i$&+15.6$i$&+13.9$i$&+12.5$i$&+11.3$i$\\
		\hline
	    $\sigma^{q}$  &2.7  &11.6&19.1 &24.2 &27.2&28.5&28.8  \\
	   \hline
	   $\sigma^{g}$  &2.0&5.2&5.7&5.0&4.1&3.2&2.5 \\
	   \hline
	    $\sigma$  &8.0&28.0&39.7&44.9&46.1&45.2&43.0 \\
		\hline\hline
	\end{tabular}
	\begin{tabular}{c c c c c c c c c }
		\hline\hline
		$\sqrt{s}$  & 4.8 & 4.9 & 5& 6& 7 &8&9 &10 \\
		\hline
		\multirow{2}*{$F(s)$} &13.6&12.0&10.6&4.0&2.0&1.1&0.7&0.5  \\
		&+10.2$i$ &+9.3$i$&+8.5$i$ &+3.9$i$ &+2.1$i$ &+1.3$i$&+0.8$i$&+0.6$i$ \\
		\hline
		$\sigma^q$& 28.2 & 27.1 & 25.7& 11.4&4.7&2.0 &0.9&0.5 \\
		\hline
		$\sigma^{g}$  &1.9&1.4&1.1&0.09&0.01&0.002&0.0005&0.0001 \\
		\hline
		$\sigma$&40.2&37.2&34.2&12.9&5.0&2.1&1.0&0.5 \\
		\hline\hline
	\end{tabular}
\end{table}

From the tables, several key observations can be made. Firstly, the cross sections from the gluonium component are significantly smaller than those from the quark-antiquark component for both processes $e^+ e^- \rightarrow J/\psi \eta$ and $e^+ e^- \rightarrow J/\psi \eta^\prime$. We find that the gluonium  contribuion is suppressed by a factor of  $M_{\eta/\eta^\prime}^2/s$. This suppression arises because the gluonium component contributions predominantly stem from two on-shell gluons at leading twist, and the corresponding matrix elements are proportional to the squared invariant mass of the two-gluon state, i.e., $M_{\eta/\eta^\prime}^2$~\cite{Baier:1981pm}.
Furthermore, from the perspective of the QCD evolution of the gluon distribution amplitude, gluonium contributions are expected to remain small. This is because the gluonium content is generally considered a higher-order effect~\cite{Kroll:2002nt,Ali:2003kg,Ali:2000ci}. In fact, this suppression is also observed in other processes, such as $J/\psi\to\eta^\prime\gamma$ and $J/\psi\to\eta^\prime e^{+}e^{-}$, where the gluonium contributions are similarly negligible~\cite{He:2019mpy,He:2020jvj}.
Notably, the relative importance of the gluonium component in $e^+ e^- \rightarrow J/\psi \eta$  is significantly smaller than that in $e^+ e^- \rightarrow J/\psi \eta^\prime$. This finding supports the indication from QCD sum rules~\cite{DeFazio:2000my}. 

Secondly, the cross section for $\eta$ production is considerably smaller than that for $\eta^\prime$ at the same CM energy.  This difference is primarily attributed to the distinct decay constants involved in each process. Specifically, the factor $\sqrt{2}f_q\cos\phi-f_s\sin\phi$ appears in the form factor $F(s)$ for the $\eta$ process, while the factor $\sqrt{2}f_q\cos\phi+f_s\sin\phi$ appears for the $\eta^\prime$ process. The former is significantly smaller than the latter, leading to the observed discrepancy in the cross sections. 

Thirdly, the cross sections for both $\eta$ and $\eta^\prime$ exhibit a characteristic behavior, increasing first and then decreasing with the rise in CM energy. This behavior is driven by the interplay between the phase space factor and the form factor $F(s)$. Specifically, $F(s)$ typically decreases with increasing CM energy, while the phase space factor increases monotonically within the considered energy range. At some intermediate energy, the product of the phase space factor and $F(s)^2$ reaches its maximal value. 

Lastly,  the cross sections for both $\eta$ and $\eta^\prime$ are significantly smaller than the experimental measurements. This discrepancy makes it challenging to explain the experimental data based solely on non-resonant contributions.

We must acknowledge that our theoretical predictions based on factorization are not reliable when $\sqrt{s}$ is closed to the threshold of $J/\psi+\eta/\eta^\prime$ (for example, $\sqrt{s}<4$ GeV). Nevertheless, these predictions still provide meaningful  
theoretical estimations for the cross sections. 

Finally, we aim to compare our calculations with those presented in Ref.~\cite{Qiao:2014pfa}. Although we have adopted a similar theoretical framework to compute the non-resonant contributions, our results differ significantly from those in Ref.~\cite{Qiao:2014pfa}, where the cross sections are approximately within an order of magnitude of the experimental data. It is intriguing to examine whether the differences in our respective treatments are responsible for the significant discrepancies observed in the predicted cross sections.
First, instead of introducing a mixing between the quark-antiquark component and gluonium component in $|\eta^\prime\rangle$  as done in Ref.~\cite{Qiao:2014pfa}, the gluonium contribution is induced through the matrix element $\langle \eta/\eta^\prime|AA|0\rangle$, which has a non-vanishing contribution due to the $U_A(1)$ anomaly~\cite{Ball:2007hb}. The decay constants for the gluonium component are related to those for the quark-antiquark component dynamically. Second, the authors in Ref.~\cite{Qiao:2014pfa} do not include the tree-level Feynman diagrams for the gluonium component of $\eta^\prime$ production (see Fig.~\ref{fig-feynman-diagram-gluon}). Instead, they consider glounium production through loop-induced diagrams, whose amplitudes are expected to be suppressed by $\mathcal{O}(\alpha_s)$ relative to the corresponding amplitudes in our work. However, since the gluonium component is already significantly suppressed compared to the quark-antiquark component, as also observed in Ref.~\cite{Qiao:2014pfa}, this difference is unlikely to have a major impact on the cross sections. Third, we use different input parameters for the LC distribution functions, charm quark mass, and mixing angles compared to those in Ref.~\cite{Qiao:2014pfa}. While these differences could contribute to some variation in the results, they are not expected to account for the order-of-magnitude discrepancy observed in the cross sections. 
The exact reason for this discrepancy remains unclear and warrants further investigation.

\subsection{A coherent sum of the resonant and non-resonant contributions}

We now specify the input parameters used to calculate the form factor $F(s)$ from intermediate resonant states. In our analysis, we include all eight vector meson resonances $V$, namely $\psi(3686)$, $\psi(3770)$, $\psi(4040)$, $\psi(4160)$, $\psi(4230)$, $\psi(4360)$, $\psi(4415)$, and $\psi(4660)$. Note the $J/\psi$ is far from the threshold of $J/\psi+\eta$ and thus contributes negligibly. Therefore, we exclude it from our analysis. 

We take the masses and total widths of $V$ from the PDG~\cite{ParticleDataGroup:2024cfk}. For $\psi(3686)$, $\psi(3770)$, $\psi(4040)$, $\psi(4160)$, and $\psi(4415)$,
their hadronic transitions to $J/\psi\eta$ have been well-studied using a model that describes the creation of a light meson in heavy quarkonium transitions~\cite{Anwar:2016mxo}. Therefore, we adopt the partial decay width for $V\to J/\psi\eta$ from Ref.~\cite{Anwar:2016mxo}. Meanwhile, the their branching fractions to $e^+e^-$  are taken from the PDG. 

For the $\psi(4230)$, $\psi(4360)$, and $\psi(4660)$ mesons, 
reliable studies on the branching ratio ${\rm Br}(V\to e^+e^-)$ and the partial decay width $\Gamma(V\to J/\psi \eta)$ are lacking in the literature. However, the combined factor $\Gamma(V\to e^+e^-){\rm Br}(V\to \jpsi H)$ is available from the PDG for these particles. Consequently, the product $\Gamma(V\to e^+e^-)\times\Gamma(V\to J/\psi H)$ can be readily obtained, which is essential for our theoretical predictions. Thus, we will use this product from the PDG for the  $\psi(4230)$, $\psi(4360)$ and $\psi(4660)$ mesons~\footnote{It is worth noting that the values of $\Gamma(V\to e^+e^-){\rm Br}(V\to \jpsi H)$ for $\psi(4230)$ and $\psi(4360)$ listed in the PDG~\cite{ParticleDataGroup:2024cfk} are derived from fits performed by the BESIII collaboration~\cite{BESIII:2020bgb}. In Ref.~\cite{BESIII:2020bgb}, three distinct solutions were identified, each with comparable fit quality and identical masses and widths for the structures around $\psi(4230)$ and $\psi(4360)$. However, the decay width for $\psi(4230)$ reported in Ref.~\cite{BESIII:2020bgb} differs significantly from that adopted in the PDG. Upon further investigation, we found that the value of $\Gamma(V\to e^+e^-){\rm Br}(V\to \jpsi H)$ from "Solution 2" in Ref.~\cite{BESIII:2020bgb}, in conjunction with the decay width provided by the PDG, can most closely reproduce the experimental resonant peak around $\psi(4230)$  in the invariant mass distribution of $e^+e^-\to J/\psi\eta$. Therefore, we have chosen to use the value of $\Gamma(V\to e^+e^-){\rm Br}(V\to \jpsi H)$ from "Solution 2" in Ref.~\cite{BESIII:2020bgb} for our calculations.}. 

Using above parameters, the effective coupling constants $g_{\gamma V}$ and $g_{VV^\prime H}$ or their product, can be determined using Eqs. (\ref{width-Vee}) and (\ref{gvvp-m}), respectively. All the input parameters utilized in our analysis are summarized 
in Tables~\ref{tab:RE} and ~\ref{tab:RE-1}.  Additionally, we set the parameter $\beta$ in (\ref{FF2}) to $400$ MeV.  

\begin{table}[!htbp]\small
	\caption{The masses, total decay widths, and ${\rm Br}(V
 \to e^+e^-)$ are taken from PDG~\cite{ParticleDataGroup:2024cfk}, and the partial decay width $\Gamma(V
 \to J/\psi \eta)$ is taken from Ref.~\cite{Anwar:2016mxo}. The central values of the coupling constants $g_{\gamma V}$ and $g_{VV^\prime \eta}$ are determined using (\ref{width-Vee}) and (\ref{gvvp-m}), with $\beta=400$ MeV.}
	\label{tab:RE}
	\centering
	\setlength{\tabcolsep}{6pt}
	\renewcommand{\arraystretch}{1.6}
	\begin{tabular}{ c|c|c|c|c|c|c }
		\hline\hline
		$V$ & $M_V $& $\Gamma_{\rm{total}}(\rm{MeV})$&Br$(V\to e^+ e^-)$&$g_{\gamma V}$  & $\Gamma(V \to J/\psi\eta)$(MeV)~\cite{Anwar:2016mxo} & $g_{VV^\prime \eta}$\\
		\hline
		\multirow{2}*{$\psi({3686})$}& \multirow{2}*{3.686} & 0.293 & \multirow{2}*{$7.94^{+0.22}_{-0.22} \times 10^{-3} $} & \multirow{2}*{0.016} & \multirow{2}*{0.01} & \multirow{2}*{0.227}\\
		& & $\pm 0.009 $& &   & & \\
		\hline
		$\psi(3770)$& $3.7737$ &  $ 27.2 \pm 1 $&$9.6^{+0.7}_{-0.7} \times 10^{-6} $ &0.0051 &0.025 & 0.157\\
		\hline
		$\psi(4040)$& $4.040 $ & $ 84 \pm 12 $&$1.02^{+0.17}_{-0.17} \times 10^{-5} $ &0.0090  &0.347& 0.294 \\
		\hline
		$\psi(4160)$& $4.191  $ &  $ 69 \pm 10 $&$6.9^{+3.3}_{-3.3} \times 10^{-6} $ & 0.0066&0.204  &0.200\\
		\hline
		$\psi(4415)$& $4.415$ &$ 110 \pm 22 $& $5.3^{+1.2}_{-1.2} \times 10^{-6} $  & 0.0071  & 0.425 &0.277\\
		\hline\hline
	\end{tabular}
 \end{table}

 \begin{table}[!htbp]
	\caption{The masses, total decay widths, and the product of $\Gamma(V
 \to e^+e^-)$ and ${\rm Br}(V
 \to J/\psi \eta)$ are taken from PDG~\cite{ParticleDataGroup:2024cfk}.
 The central values of the product of the coupling constants $g_{\gamma V}$ and $g_{VV^\prime \eta}$ are determined using (\ref{width-Vee}) and (\ref{gvvp-m}), with $\beta=400$ MeV.}
     \centering
     \label{tab:RE-1}
	\begin{tabular}{ c|c|c|c|c }
	\hline\hline
	$V$ & $M_V$& $\Gamma_{\rm total}(\rm {MeV})$ & $\Gamma(V\to e^+e^-){\rm Br}(V\to \jpsi \eta)$(GeV) & $g_{\gamma V}\times g_{VV^\prime \eta}$ \\
	\hline
	$\psi(4230)$&$4.2221 $ &  $ 49 \pm 7 $ & $4.8^{+1.0}_{-1.0}\times 10^{-9}$  &$3.08 \times 10^{-8}$\\
	\hline
	$\psi(4360)$& $4.374 $ &$ 118 \pm 12 $  &  $1.7^{+1.1}_{-1.1}\times 10^{-9}$ &$2.51\times10^{-8}$\\
	\hline
	$\psi(4660)$& $4.641$ &$ 73^{+13}_{-11} $&$< 0.94 \times 10^{-9} $ &$8.95\times10^{-9}$\\
	\hline\hline
	\end{tabular}
\end{table}

We begin by computing the form factor $F(s)$ which receives contributions from both non-resonant and resonant parts. The form factor
$F(s)$ can be expressed as
\begin{align}\label{vmd-amp-total}
	\notag
	F(s) =&F^{\rm nonres}(s)+F^{\rm VMD}_{\psi_{(3686)}}(s) +e^{i\phi_1} F^{\rm VMD}_{\psi_{(3770)}}(s) +e^{i\phi_2}F^{\rm VMD}_{\psi(4040)}(s) +e^{i\phi_3}F^{\rm VMD}_{\psi(4160)}(s) \\
	&+e^{i\phi_4}F^{\rm VMD}_{\psi(4230)}(s) +e^{i\phi_5}F^{\rm VMD}_{\psi(4360)}(s) +e^{i\phi_6}F^{\rm VMD}_{\psi(4415)}(s) +e^{i\phi_7}F^{\rm VMD}_{\psi(4660)}(s)
\end{align}
where $\phi_i(i=1,2,\cdots,7)$ denote the relative phase angles between different resonances, $F^{\rm nonres}(s)$  represents the contribution from the non-resonant part. Given that $F^{\rm nonres}(s)$ is negligibly small, we simply take it in phase with the $\psi(3686)$. Since it is challenging to constrain these relative phases directly, we treat them as free parameters. Their values will be determined through a least-$\chi^2$  fit to the invariant mass distribution of the process $e^+e^-\to J/\psi \eta$.

Using Eqs. (\ref{cross-section-re}) and (\ref{vmd-amp-total}), we can readily obtain the theoretical predictions for the cross section of  $e^+e^-\to J/\psi\eta$. The phase angles $\phi_i$ are determined through a least-$\chi^2$ fit to the experimental data from the BESIII collaboration~\cite{BESIII:2023tll} and the Belle collaboration~\cite{Wang:2012bgc}. The values of these angles (in units of radians) are as follows:
\\
For fitting the BESIII data~\cite{BESIII:2023tll} (Set I):
\bqa\label{fit-1}
\phi_1&=&-0.92,\,\phi_2=1.84,\,\phi_3=-0.02,\,\phi_4=-0.58,\nn\\
\phi_5&=&-0.99,\,\phi_6=-2.14,\,\phi_7=-1.78.
\eqa
For fitting the Belle data~\cite{Wang:2012bgc} (Set II):
\bqa\label{fit-2}
\phi_1&=&-0.19,\,\phi_2=3.18,\,\phi_3=3.40,\,\phi_4=-0.07,\nn\\
\phi_5&=&-1.55,\,\phi_6=-2.97,\,\phi_7=-3.63.
\eqa
For fitting both experiments simultaneously (Set III):
\bqa\label{fit-3}
\phi_1&=&-0.86,\,\phi_2=1.91,\,\phi_3=0.01,\,\phi_4=-0.56,\nn\\
\phi_5&=&-0.96,\,\phi_6=-2.08,\,\phi_7=-2.10.
\eqa

\begin{figure}[htbp]
	\centering
	\includegraphics[width=0.85\textwidth]{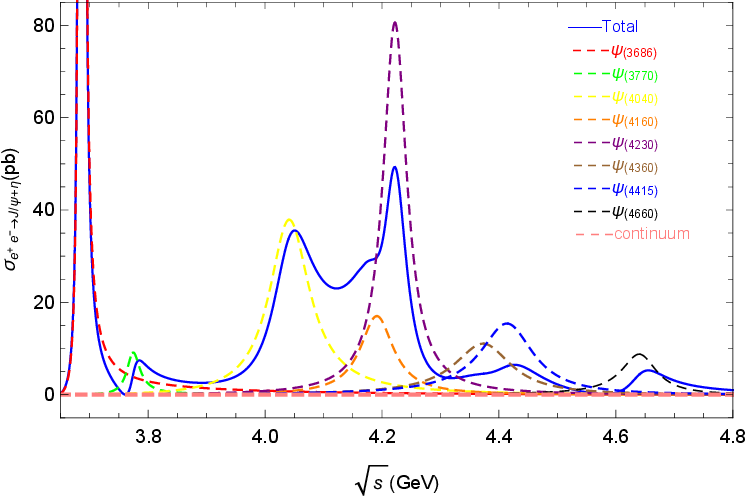}
	\caption{The cross sections from each individual resonance state and the total cross section are shown as a function of  $\sqrt{s}$.
The pink dashed line indicates the contribution from the non-resonant continuum, while the dashed lines in various colors correspond to the contributions from different resonance states, respectively. The blue solid line represents the total cross section using the phase angles from Eq. (\ref{fit-3}) (Set III).
		\label{VMD-cs-part}}
\end{figure}

By adopting the set of phase angles given in Eq. (\ref{fit-3}), we present the theoretical predictions in Fig.~\ref{VMD-cs-part} for the differential cross section of $e^+e^-\to J/\psi\eta$ as a function of $\sqrt{s}$. The contributions from the non-resonant continuum, individual resonant states, and their combined interference are shown separately. From the figure, it is evident that there is significant interference among the contributions from the resonant states. Specifically, the resonance peaks around $4230$ MeV and $4415$ MeV  are notably suppressed due to this interference.

\begin{figure}[H]
	\centering
    \includegraphics[width=0.6\linewidth]{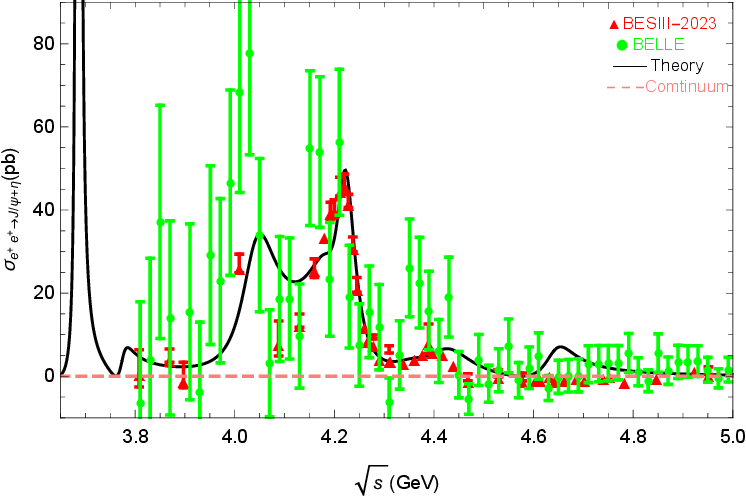}\\
	\caption{The black solid line represents the distribution of the total cross section using phase angles from Set I. The red triangles with error bars denote the measurements from BESIII~\cite{BESIII:2023tll}, the green dots with error bars represent the measurements from Belle~\cite{Wang:2012bgc}, and the pink dashed line indicates the contribution from the non-resonant continuum.}
	\label{fig:compare-exp-SetI}
\end{figure}

\begin{figure}[H]
	\centering
    \includegraphics[width=0.6\linewidth]{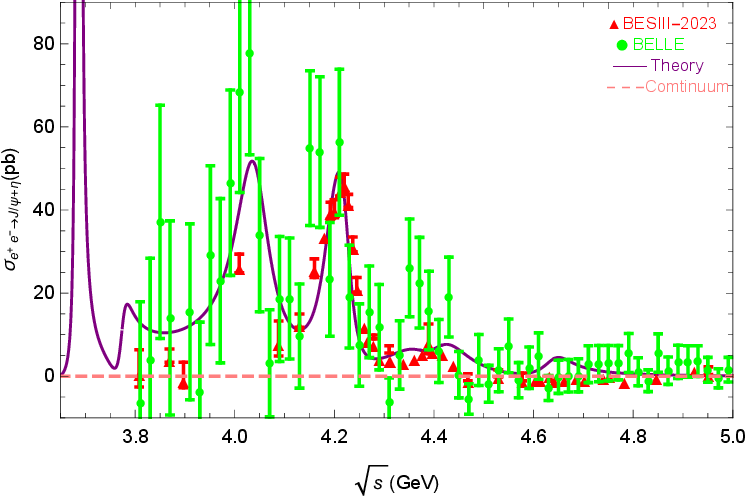}\\
	\caption{The purple The black solid line represents the distribution of the total cross section using phase angles from Set II. The red triangles with error bars denote the measurements from BESIII~\cite{BESIII:2023tll}, the green dots with error bars represent the measurements from Belle~\cite{Wang:2012bgc}, and the pink dashed line indicates the contribution from the non-resonant continuum.}
	\label{fig:compare-exp-SetII}
\end{figure}

\begin{figure}[H]
	\centering
    \includegraphics[width=0.7\linewidth]{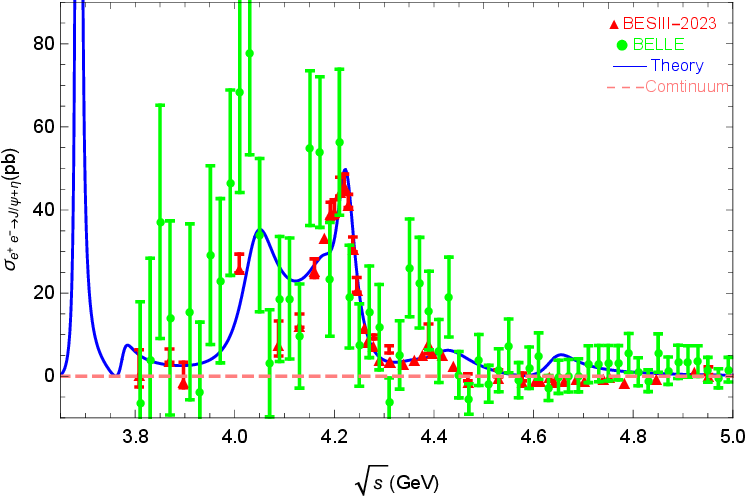}\\
	\caption{The blue solid line represents the distribution of the total cross section using phase angles from Set III. The red triangles with error bars denote the measurements from BESIII~\cite{BESIII:2023tll}, the green dots with error bars represent the measurements from Belle~\cite{Wang:2012bgc}, and the pink dashed line indicates the contribution from the non-resonant continuum.}
	\label{fig:compare-exp-SetIII}
\end{figure}

We proceed to compare our theoretical predictions with the experimental data for the invariant mass distributions of the process $e^+e^-\to J/\psi\eta$. The comparisons between our theoretical predictions and the BESIII data~\cite{BESIII:2023tll} as well as the Belle data~\cite{Wang:2012bgc} are presented in Figs.~\ref{fig:compare-exp-SetI}–\ref{fig:compare-exp-SetIII}. Specifically, the phase angles from Set I, Set II, and Set III are used in Fig.~\ref{fig:compare-exp-SetI}, Fig.~\ref{fig:compare-exp-SetII}, and Fig.~\ref{fig:compare-exp-SetIII}, respectively. 

From the figures, several key observations can be made. The theoretical predictions using the phase angles from Set I and Set III both exhibit a resonant peak around $4040$ MeV, but they fail to accurately describe the peak structure observed around $4040$ MeV in the Belle data~\cite{Wang:2012bgc}. In fact, we have found that the theoretical cross section resulting solely from the resonance at 4040 MeV is significantly smaller than that observed in the Belle data.

Furthermore, the peak structure around 4360 MeV in the Belle data cannot be accurately explained by the theoretical predictions from any of the three sets of phase angles. This discrepancy may arise from significant interference between the resonances $\psi(4360)$ and $\psi(4415)$ in our theoretical predictions. Additionally, a prominent peak structure around 4660 MeV is observed in our theoretical predictions but is absent in the BESIII data~\cite{BESIII:2023tll}. We anticipate that future experiments with larger datasets could potentially search for these structures.

Finally, we acknowledge that our theoretical predictions are based on the central values of the input parameters, and uncertainties in these parameters have not been considered in our analysis. We expect that these uncertainties could influence our theoretical predictions and the comparison between theory and experiment.  

Comparing the cross sections of $e^+e^- \to J/\psi \eta$ and $e^+e^- \to  J/\psi \eta^\prime$ is of particular interest. However, predicting the cross sections for $e^+e^- \to  J/\psi \eta^\prime$ within the VMD framework remains challenging due to the absence of both theoretical calculations and experimental measurements for the hadronic transition process $\psi\to J/\psi\eta^\prime$. As a result, we directly utilize the measured cross sections for $e^+e^- \to  J/\psi \eta^\prime$  from the BESIII collaboration~\cite{BESIII:2019nmu}, while relying on our theoretical predictions for the cross sections of $e^+e^- \to  J/\psi \eta$.  The values for both processes are summarized in Table~\ref{tab:eta-etap-compare} across a range of CM energies. In this table, the cross sections for $e^+e^- \to  J/\psi \eta$ are calculated using three distinct sets of phase angles. It is clear from the comparison that the cross sections for $e^+e^- \to  J/\psi \eta^\prime$ are significantly smaller than those for $e^+e^- \to  J/\psi \eta$. 
\begin{table}[!htbp]\small
	\caption{ $\sigma_\eta$ and $\sigma_{\eta^\prime}$  (in units of pb) denote the cross section for the process $e^+e^-\to J/\psi \eta $ and  $e^+e^-\to J/\psi \eta^{\prime}$, respectively, in the CM energy range from 4.178 to 4.600 GeV. The superscript I, II, and III in $\sigma_{\eta}$ indicate the cross sections calculated using the relative phase angles from Set I, Set II, and Set III, respectively. $\sigma_{\eta^{\prime}}$ is taken from the BESIII collaboration~\cite{BESIII:2019nmu}.}
	\label{tab:eta-etap-compare}
	\centering
	\setlength{\tabcolsep}{8pt}
	\renewcommand{\arraystretch}{1.6}
	\begin{tabular}{c c c c c c c c}
		\hline\hline
		$\sqrt{s}$ & 4.178 & 4.189 & 4.199 & 4.209 & 4.219 &4.226&4.236 \\
		\hline
		$\sigma_{\eta^{\prime}}$ &$2.43^{+0.34}_{-0.34}$ & $2.21^{+0.75}_{-0.75}$  &$2.87^{+0.85}_{-0.85}$ &$2.68^{+0.78}_{-0.78}$ &$2.46^{+0.77}_{-0.77}$ &$3.63^{+0.66}_{-0.66}$&$2.85^{+0.85}_{-0.85}$\\
		\hline
		$\sigma_{\eta}^{\rm{I}}$ &29.2& 29.8& 32.6& 40.6&49.0&48.0&36.4 \\
		\hline
		$\sigma_{\eta}^{\rm{II}}$ &27.0& 37.0& 45.2&49.1&44.0&34.3&19.7\\
		\hline
		$\sigma_{\eta}^{\rm{III}}$  &29.2 &29.7  &32.4&40.5&49.1&48.2&36.7\\
		\hline\hline
	\end{tabular}
	\begin{tabular}{c c c c c c c c}
		\hline\hline
		$\sqrt{s}$ & 4.244&4.258 &4.267&4.278 &4.358 &4.416&4.600 \\
		\hline
    $\sigma_{\eta^{\prime}}$&$3.81^{+0.93}_{-0.93}$&$3.41^{+0.71}_{-0.71}$&$2.83^{+0.73}_{-0.73}$ &$0.45^{+0.45}_{-0.45} $ &$ 0.21^{+0.21}_{-0.21}$ &$ 1.18^{+0.38}_{-0.38}$ &$0.21^{+0.31}_{-0.31} $ \\
		\hline
		$\sigma_{\eta}^{\rm{I}}$ &26.0& 13.8&9.6& 6.5&4.3&6.4&0.6 \\
		\hline
		$\sigma_{\eta}^{\rm{II}}$ &12.1& 6.2& 4.8 & 4.2 &6.5&7.5&1.0\\
		\hline
		$\sigma_{\eta}^{\rm{III}}$  &26.2  &14.0 &9.7&6.5&4.0&6.0&0.4\\
		\hline\hline
	\end{tabular}
\end{table}

It is worth noting that if future theoretical calculations or experimental measurements become available for the process $\psi\to J/\psi\eta^\prime$, it would be feasible to predict the cross sections for $e^+e^- \to  J/\psi \eta^\prime$ using the framework adopted in this work.

\section{summary~\label{sec-summary}}
The cross sections for $e^+e^-\to \jpsi \eta$ and $e^+e^-\to \jpsi \eta^\prime$ are calculated using the framework of NRQCD combined with LC factorization. The predicted cross sections are on the order of several femtobarns for  
$e^+e^-\to \jpsi \eta^\prime$, and less than 1 fb for $e^+e^-\to \jpsi \eta$.  These values are several orders of magnitude smaller than the experimental measurements reported by the BESIII and Belle collaborations. This significant discrepancy suggests that resonant contributions dominate the cross sections at the CM energies probed by these experiments.

To account for these resonant contributions, we employ the VMD model. This requires determining two effective coupling constants: one between the photon and the resonance, and another between the resonance and $J/\psi\eta$. These coupling constants are extracted from the decay widths $\Gamma(\psi\to e^+e^-)$ and 
$\Gamma(\psi\to J/\psi\eta)$, which are either obtained from the PDG or computed theoretically.

We treat the predictions from the factorization calculation as the continuum (non-resonant) contribution and calculate the cross section for $e^+e^-\to \jpsi \eta$ by summing coherently the contributions from various resonances and the continuum. The relative phase angles between different resonances in the amplitude are determined through a least-$\chi^2$ fit to the experimental data. Upon comparing our theoretical predictions with the experimental data, we find that our model poorly describes the peak structure around $4360$ MeV observed in the Belle data. Furthermore, our theoretical calculation predicts a resonance peak around 4660 MeV, which is not observed in the BESIII measurements. Undoubtedly, these discrepancies highlight the need for further theoretical investigation and more precise experimental measurements.

Additionally, we observe that the theoretical predictions for the cross section of $e^+e^-\to \jpsi \eta$ are significantly larger—by an order of magnitude—than those for $e^+e^-\to \jpsi \eta^\prime$ measured experimentally at most CM energies. This finding is consistent with experimental observations.

\section*{Acknowledgments}
The work of C.-M. G., W.-L. S., and M.-Z. Z. is supported by the
National Natural Science Foundation of China under Grants 
No. 12375079, and the Natural Science
Foundation of ChongQing under Grant No. CSTB2023
NSCQ-MSX0132. The work of J.-K. H. is supported by the National Natural Science Foundation of China under Grant No. 12305086, and the Open Fund of the Key Laboratory of Quark and Lepton Physics (MOE) under Grant No. QLPL2024P01.

\end{document}